\documentstyle[epsf,twocolumn]{article}

\begin{document}
\renewcommand{\textfraction}{0.1}
\renewcommand{\topfraction}{0.8}
\rule[-8mm]{0mm}{8mm}
\begin{minipage}[t]{16cm}
{\large \bf Theory of Magnetic Short--Range Order 
for High--$\rm \bf T_c$ Superconductors\\[4mm]}
U.~Trapper$^a$, D. Ihle$^a$ and H.~Fehske$^b$\\[3mm]
$^a${Institut f\"ur Theoretische Physik, Universt\"at Leipzig, 
D--04109 Leipzig, Germany}\\
$^b${Physikalisches Institut, Universt\"at Bayreuth, 
D--95440 Bayreuth, Germany}\\[4.5mm]
\hspace*{0.5cm}A theory of magnetic short--range order for high--$T_c$
cuprates is presented on the basis of the one--band 
$t$--$t^{\prime}$--Hubbard model combining the four--field
slave--boson functional integral 
technique with the Bethe cluster method. The ground--state 
phase diagram evaluated self--consistently at the saddle--point 
and pair--approximation levels shows the experimentally
observed suppression of magnetic long--range order  
in the favour of a paraphase with antiferromagnetic 
short--range order. In this phase the uniform static spin susceptibility
consists of interrelated itinerant and local parts and increases upon doping 
up to the transition to the Pauli paraphase. Using realistic values
of the Hubbard interaction we obtain the cusp position and the 
doping dependence of the zero--temperature susceptibility in reasonable
agreement with experiments on $\rm La_{2-\delta}Sr_{\delta}CuO_4$.
\end{minipage}\\[4.5mm]
\normalsize
The spin correlations in high-$T_c$ superconductors probed
by neutron and NMR experiments and by the spin susceptibility in the
paraphase (showing, for $\rm La_{2-\delta}Sr_{\delta}CuO_4$ (LSCO), a maximum
in the doping and temperature dependences~\cite{Toea89_Jo89}) are
believed to be caused by a strong Coulomb interaction in the $\rm
CuO_2$ planes which may result in a considerable antiferromagnetic
short-range order (SRO). In our previous work \cite{TIF95} the role 
played by SRO in explaining the unconventional magnetic properties of
the cuprates was investigated within the 2D one--band Hubbard model
based on the scalar four--field slave--boson (SB) approach~\cite{KR86_Ha89}.

In this paper the SRO concept is futher elaborated, where attention is
paid to band--structure effects on the stability of magnetic SRO
versus long--range order (LRO) and on the doping dependence of the
spin susceptibility. In the SB representation the Hubbard model can be 
expressed as~\cite{KR86_Ha89}
\begin{equation}
{\cal H} = \sum_{ij\sigma}
t_{ij}^{ }
z_{i\sigma}^\dagger f_{i\sigma}^\dagger f_{j\sigma}^{ } z_{j\sigma}^{ } 
\nonumber + U\,\sum_i d_i^\dagger d_i^{ }\;.
\end{equation}
The functional
integral for the partition function is calculated in the radial gauge
and the static approximation for the SB fields $p_{i\sigma}$, $d_i$,
and the Lagrange multiplier field $\lambda_{i\sigma}^{(2)}$ (enforcing
the constraint $f_{i\sigma}^\dagger f_{i\sigma}^{ } =
 p_{i\sigma}^{\dagger}p_{i\sigma}^{ } + d_i^{\dagger} d_i^{ }$), where
the $e_i$ fields are eliminated by the saddle--point approximation for
$\lambda_{i}^{(1)}$ (guaranteeing $e_i^{\dagger} e_i^{ } +
d_i^{\dagger} d_i^{ } + 
\sum_\sigma p_{i\sigma}^{\dagger}p_{i\sigma}^{ } 
= 1$)~\cite{KR86_Ha89}.
\begin{figure}[t]
\unitlength1mm
\begin{picture}(70,72)
\end{picture}
\end{figure}
Transforming away the SB fluctuations  in the transfer term
(proportional to the band narrowing $z_{i\sigma}^{ }$), we get the
free--energy functional
\begin{equation}
\label{FI}
{\mit \Psi} = \sum_i\left(U d_i^* d_i^{ } - n_i \nu_i + m_i
\xi_i\right)-\mbox{Tr} 
\ln{\left[-\hat{G}_{ }^{-1}\right]} \;, 
\end{equation}
\begin{equation}
\label{greens}
 \hat{G}_{ij\sigma}^{-1} = 
\frac{\omega -\nu_i+ \sigma (\xi_i+h)}
        {\left|z_{i\sigma}^{ }\right|^2} \delta_{ij}-t_{ij} \;, 
\end{equation}
where $m_i=\sum_{\sigma} \sigma p_{i\sigma}^2$ and  $n_i=\sum_{\sigma}
p_{i\sigma}^2$ are the SB representations of the local magnetization
and particle number, respectively, $\xi_i=\sum_{\sigma} \sigma
\lambda_{i\sigma}^{(2)}$ is the internal magnetic field, and 
$\nu_i=\sum_{\sigma} \lambda_{i\sigma}^{(2)}$. In (\ref{greens}), 
$h$ denotes the uniform external magnetic field. 

We incorporate the SRO beyond the PM saddle point by an expansion in
terms of the local perturbation
\begin{equation}
V_{i\sigma}(\omega)=-\hat{G}_{ij\sigma}^{-1} + \hat{G}_{ij\sigma}^{o-1}\,.
\end{equation}
$\hat{G}_{ij\sigma}^{o}={\rm FT}\{[(\omega-\nu^o +\sigma(\xi^o+h))/(z_{\sigma}^{o})_{ }^2 -
\varepsilon_{\vec{k}}]^{-1}\}$ is the PM Green
propagator (the superscript ``$o$'' denotes the {\it uniform} paramagnetic
(PM) saddle point), where 
\begin{equation}
\label{disp}
\varepsilon_{\vec{k}}=-2t(\cos k_x +\cos k_y )
-4t^{\prime}\cos k_x \cos k_y
\end{equation}
is the 2D tight--binding dispersion taking into account the transfer
integral between nearest ($t$) and next--nearest ($t^{\prime}$)
neighbours. 
Now we
treat the fluctuations of $m_i=\bar{m}_i s_i$, $\xi_i=\bar{\xi}_i s_i$
($s_i=\pm$) and of the charge degrees of freedom by the ansatz $b_i
\to b_{s_i}$ with $b \in \{ \bar{m}, \bar{\xi},n,\nu,d=d^*\}$.
Transforming the functional (\ref{FI}) to an effective Ising
model in the nearest--neighbour pair ($\langle ij \rangle$)
approximation~\cite{Ka81a} we obtain
\begin{equation}
\label{PsiIs}
{\mit \Psi}(\{s_i\}) = \bar{\mit \Psi} -\bar{h} \sum_i s_i 
- \bar{J} \sum_{\langle ij \rangle} s_i s_j\;,
\end{equation}
with
\begin{equation}
\label{JIsing}
\bar{J} = -\mbox{$\frac{1}{4}$}\sum_{\alpha,\sigma=\pm}\left(
                 {\mit \Phi}_{\alpha \alpha \sigma}
                -{\mit \Phi}_{{-\alpha} \alpha \sigma} \right) \;,
\end{equation}
where the two--site fluctuation 
contribution 
${\mit \Phi}_{\alpha \alpha^{\prime}\sigma}=\left.
{\mit \Phi}_{\langle ij\rangle \sigma}(\alpha_i^{ },\alpha_j^{ })
\right|_{{\alpha_i^{ }=\alpha}^{ }\atop {\alpha_j^{ }=\alpha^{\prime}}}$
is found to be 
\begin{eqnarray}
\label{phiaa}
\lefteqn{{\mit \Phi}_{\langle ij \rangle \sigma}^{ } = \mbox{$\frac{1}{\pi}$}
\int d\omega f(\omega - \mu) }\nonumber \\
&&\hspace*{-0.5cm}\mbox{Im}~\ln\left[1-G_{\langle ij\rangle\sigma}^{o}
T_{j\sigma}(\alpha_j^{ })G_{\langle ji\rangle\sigma}^{o}
T_{i\sigma}(\alpha_i^{ })\right],\,
\end{eqnarray}
and $T_{i\sigma}^{ }= V_{i\sigma}^{ }(1-G_{ii\sigma}^{o} 
V_{i\sigma}^{ })^{-1}$ denotes the scattering matrix.
Performing the $s_i$ sum in the partition function with
(\ref{PsiIs}) we treat the SRO in the Bethe cluster
approximation. Thereafter, we determine the saddle point for all Bose
fields $b_{\alpha} \in \{ \bar{m}_{\alpha},
\bar{\xi}_{\alpha},n_{\alpha},
\nu_{\alpha},d_{\alpha}\}$. Correspondingly, in our theory the SRO is
self--consistently described at the saddle--point and pair
approximation levels at each interaction strength $U$ and hole doping
$\delta=1-n$. 

In the $h=0$ limit ($\bar{m}_{\alpha}=\bar{m}$) we obtain two possible
paraphases ($\langle s_i \rangle=0$): (i) the paraphase without SRO (PM;
$\bar{J}=0$, $\bar{m}=0$) and (ii) the paraphase with
antiferromagnetic SRO (SRO-PM; $\bar{J}<0$, $\bar{m}>0$). Let us stress
that, at $T=0$, the SRO-PM phase must be distinguished from the phase
with antiferromagnetic LRO (denoted by AFM) having a finite sublattice
magnetization $m_A^{ }=p_{A\uparrow}^2-p_{A\downarrow}^2=-m_B^{ }$ which is
determined from the A-B saddle point \cite{DFB92} and differs from the
SRO amplitude $\bar{m}$.

In Fig.~1 the ground--state phase diagram is depicted, where only the
phases describable by the scalar SB approach are shown and the
tight--binding density of states is used (instead of a semielliptic
one taken in Ref.~\cite{TIF95}). At large enough interaction strengths
($U/t>5-6$) the antiferromagnetic LRO makes way to antiferromagnetic
SRO in a wide doping region, where we obtain an AFM 
$\protect\rightleftharpoons$ SRO--PM phase transition of first
order at $\delta_{c_1}$ and a SRO--PM $\protect\rightleftharpoons$ PM
transition of second order at $\delta_{c_2}$. For $U/t=8$ (being realistic
for the cuprates~\cite{HSSJ90}) and $t^{\prime}=0$ we get
$\delta_{c_1}=0.04$ and   $\delta_{c_2}=0.26$. The inclusion of the
$t^{\prime}$ term in (\ref{disp}) ($t^{\prime}/t=-0.16$ is realistic
for LSCO) extends the stability region of the SRO-PM phase, since
hopping processes along the lattice diagonals favour antiferromagnetic
correlations.
\begin{figure}[h]
\centerline{\mbox{\epsfxsize 7cm\epsffile{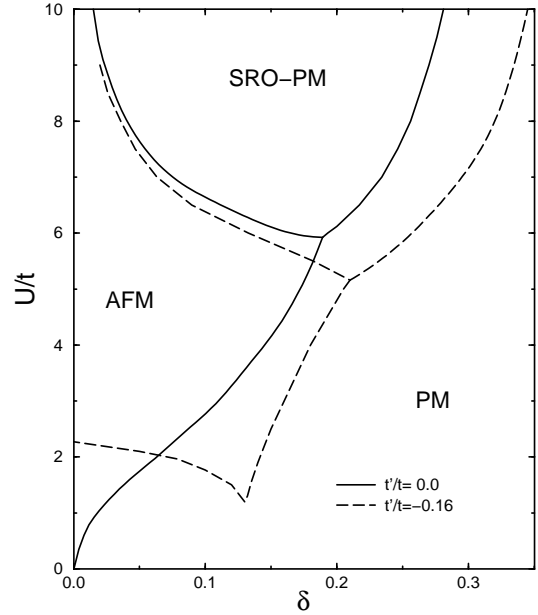}}}
\caption{Ground--state phase diagram}
\end{figure}

Within our theory, the suppression of magnetic LRO at
$\delta_{c_1}\simeq 4\%$ ($U/t \simeq 8$) observed in LSCO may be
related to the persistence of SRO in the paraphase. 
The SRO breaks down at $\delta_{c_1}\simeq 30\%$. Note that just at
this doping the superconductivity in LSCO disappears.
Correspondingly, the unconventional behavior of high--$T_c$ cuprates
may be due to the presence of magnetic SRO which we suggest from our
results to play also a role in the pairing mechanism.

The uniform static spin susceptibility $\mit \chi$ has to be
calculated according to
\begin{equation}
\label{chidef}
{\mit \chi}=\lim_{h\to 0}\,\sum_{\alpha} \left( W_\alpha \frac{d\,m_\alpha}{d\,h}
+m_\alpha \frac{d\,W_\alpha}{d\,h}\right)\;,
\end{equation}
where $m_{\alpha}=\bar{m}_{\alpha} \alpha$, $W_{\alpha}(\bar{h},
h_{ }^{*},\bar{J})$
is the probability for the Ising spin projection $\alpha$ at the 
central site of the
Bethe cluster, and $h^{*}$ is the effective Bethe field. The first term in
(\ref{chidef})  describes the change of the
magnetization amplitude with the applied magnetic field and gives
mainly the `itinerant' contribution to $\chi$. The second term
describes directional fluctuations of the local magnetizations 
(`local' contribution) and is finite only in the SRO--PM phase.
Note that the `itinerant' and `local' properties are interrelated
and determine {\it both} contributions to the spin susceptibility.

Figure~2 shows our results for the doping dependence of the
zero--temperature susceptibility $\chi(\delta)$.
In the PM phase ($\delta>\delta_{c_2}$) the SB band--renormalized Pauli
susceptibility has a pronounced doping dependence in two
dimensions which is strongly affected by $t^{\prime}$. In the 
SRO--PM phase ($\delta_{c_1}<\delta<\delta_{c_2}$),
the Pauli susceptibility is suppressed due to
the SRO--induced spin stiffness against the orientation of the local
magnetizations along the
homogeneous external field. Accordingly, at $\delta_{c_2}$ a cusp in
$\chi(\delta)$ appears, where the $U/t$ and $t^{\prime}$ dependences
of the peak--position emerge from Fig.~1. Since, for
$\delta_{c_1}<\delta<\delta_{c_2}$, 
$|\bar{J}|$ decreases with increasing $\delta$, 
the susceptibility increases upon doping.
\begin{figure}[h]
\centerline{\mbox{\epsfxsize 7cm\epsffile{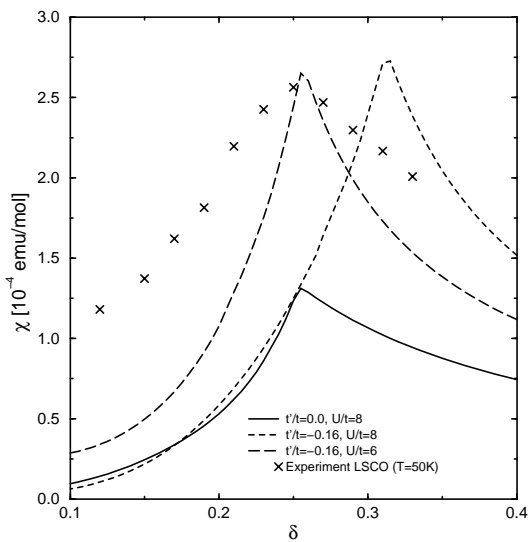}}}
\caption{Uniform static spin susceptibility at $T=0$
(taking the realistic value $t=0.3$~eV). }
\end{figure}

In Fig.~2 we have also depicted the spin contribution to the magnetic 
susceptibility of 
LSCO at 50~K
obtained from the experimental data on the total susceptibility 
by subtracting the diamagnetic core ($-9.9 \times
10^{-5}$ emu/mol) and Van Vleck ($ 2.4 \times 10^{-5}$ emu/mol) 
contributions which can be taken as 
independent of doping and temperature 
over the limited parameter region studied here~\cite{Toea89_Jo89}. The
experimentally observed pronounced maximum at a hole doping of about
25\% and the qualitative doping dependence of $\chi$ are reproduced
rather well by our theory without any fit procedure.

From our results we conclude that the concept of magnetic SRO in
strong--correlation models may play the basic role in the explanation of
many unconventional properties of high--$T_c$ compounds. As motivated
by neutron scattering probing the correlation length over
several lattice spacings, our theory should be extended by the
description of a larger than nearest--neighbour ranged SRO.

This work was performed under the auspices of
Deutsche For\-schungsgemeinschaft under project SF--HTSL--SRO.
U.T. acknowledges the hospitality at the University Bayreuth.
\bibliography{ref}
\bibliographystyle{phys}
\end{document}